\documentclass[11pt, a4paper]{article}

\usepackage[a4paper, top=2.5cm, bottom=2.5cm, left=2cm, right=2cm]{geometry}

\usepackage{amsmath} 
\usepackage{booktabs} 
\usepackage{graphicx} 
\usepackage{authblk} 
\usepackage{caption} 
\usepackage{enumitem} 
\usepackage{cprotect} 

\usepackage[colorlinks=true, linkcolor=blue, urlcolor=cyan, citecolor=green, bookmarks=true]{hyperref}

\title{Quantifying the Hemodynamic Effects of Ventricular Fibrillation using a Verified Computational Model}
\author[1]{Artemii Remizov}
\author[2]{Sergey Lapin}
\affil[1]{Lake Washington High School, Kirkland, WA}
\affil[2]{Washington State University, Everett, WA}
\date{October 17, 2025}

\begin{document}

\maketitle
\thispagestyle{empty} 
\clearpage
\setcounter{page}{1} 
\pagestyle{plain} 

\begin{abstract}
Ventricular Fibrillation (VF) is a life-threatening cardiac arrhythmia characterized by disorganized, high-frequency electrical impulses in the ventricles, leading to a profound failure of the heart's pumping function. This study details the implementation and verification of a foundational 1988 lumped-parameter model of the closed-loop cardiovascular system. The verified baseline model, representing a healthy adult, was then modified to simulate the pathophysiological conditions of VF by increasing the heart rate to 350 beats per minute and decreasing ventricular contractility by 50\%. The simulation results demonstrate a severe degradation in cardiovascular function, quantifying a 62.4\% reduction in cardiac output, from 5.88 L/min to 2.21 L/min. This quantitative analysis underscores the severity of VF and verifies the computational model as a tool for studying arrhythmias.
\par
This paper expands upon this foundational result by situating the 0D model within a state-of-the-art, multiscale research framework. We critique the limitations of the prescribed model and propose the necessary next steps for a mechanistic investigation. This includes: (1) Integrating the 0D hemodynamic core with cellular-level electrophysiological (EP) models to simulate emergent arrhythmias; (2) Coupling the system with a quantitative model of the baroreflex loop to capture the autonomic nervous system response; and (3) A pathway toward patient-specific parameterization using data assimilation. An interactive simulator based on the verified 0D model was developed to facilitate exploration and serves as a testbed for future enhancements. The source code for the interactive simulator is publicly accessible online \cite{simulatorURL}.
\end{abstract}

\newpage
\tableofcontents
\newpage

\section{Introduction}
Ventricular Fibrillation (VF) is a life-threatening arrhythmia responsible for the majority of sudden cardiac death (SCD) incidents, claiming hundreds of thousands of lives annually \cite{statVF}. The condition is characterized by disorganized, high-frequency electrical activity in the ventricles, which abolishes coordinated mechanical contraction. This acute cessation of the heart's pumping function leads to immediate, global hemodynamic compromise and, if untreated, death within minutes.

While the terminal clinical outcome is well-understood, the precise spatio-temporal dynamics that link ion-channel dysfunction to whole-organ electrical disorganization, and this electrical disorganization to mechanical failure, are profoundly complex \cite{traynova2011}. Investigating these dynamics \textit{in vivo} is exceptionally difficult. Consequently, \textit{in silico} experimentation, or computational modeling, has emerged as an indispensable tool in modern cardiology \cite{quarteroni2017, noble2012}.

This paper presents a hierarchical computational study of VF. Our work begins with the foundational principles of cardiovascular modeling: the implementation and verification of a robust, closed-loop, lumped-parameter (0D) model based on the foundational work of Avanzolini et al. \cite{avanzolini1988}. We utilize this computationally efficient model to establish a verified baseline of healthy circulation and then to quantify the global hemodynamic consequences of a \emph{prescribed} VF state. This initial simulation provides a precise, quantitative measure of circulatory compromise, highlighting a 62.4\% reduction in cardiac output.

However, prescribing an arrhythmia by simply setting a heart rate and contractility is a significant simplification. The true scientific challenge lies in modeling the \emph{emergence} of the arrhythmia itself and the body's complex physiological response. Therefore, the second, more advanced goal of this paper is to propose a research framework for integrating this verified 0D hemodynamic model into a sophisticated, multiscale simulation. We argue that the 0D model is not an end, but a vital, computationally tractable component for next-generation research. We outline a clear path forward, detailing the necessary coupling with advanced models of both cardiac electrophysiology (EP) and the Autonomic Nervous System (ANS) to create a simulator capable of bridging the gap from ion channel to whole-body circulation. Finally, we describe an interactive simulator developed based on the verified 0D model, whose source code is publicly accessible online \cite{simulatorURL}, serving as both an educational tool and a platform for demonstrating the model's capabilities and future potential.

\section{Overview of Cardiovascular Modeling Frameworks}
Computational modeling of the cardiovascular system is a mature field, characterized by a hierarchy of models that balance physiological detail with computational cost \cite{smith2011}. Understanding this hierarchy is essential to contextualize the 0D model and the advanced framework we propose.

\subsection{Lumped-Parameter (0D) Models}
Lumped-parameter (0D) models are the most computationally efficient representation of the cardiovascular system. They are based on an electrical-hydraulic analogy, where the system is simplified into a network of compartments (e.g., left ventricle, systemic arteries) \cite{shi2011}. In this analogy, pressure ($P$) is analogous to voltage, flow ($Q$) to current, vascular resistance ($R$) to electrical resistance, and vascular compliance ($C$) to capacitance.

The pioneering work of Guyton, Sagawa, and others established the core principles of this approach \cite{sagawa1974, guyton1972}. Models like the one by Avanzolini et al. \cite{avanzolini1988} expanded this into a comprehensive closed-loop system, including all four heart chambers, valves, and both systemic and pulmonary circulations. The primary strength of 0D models is their ability to simulate global hemodynamic behavior (e.g., cardiac output, mean arterial pressure) over many heartbeats in real-time. Their main limitation is the complete absence of spatial information. They cannot describe pressure waves, regional blood flow, or the propagation of electrical signals.

\subsection{One-Dimensional (1D) Models}
One-dimensional (1D) models represent the major arteries as a network of elastic tubes. By solving the 1D, cross-section-averaged Navier-Stokes equations, these models can accurately simulate the propagation of pressure and flow waves throughout the arterial tree \cite{quarteroni2000}. This is critical for studying phenomena like wave reflection, which significantly impacts central aortic pressure and left ventricular load. 1D models are often coupled with 0D models, which are used to represent the heart at the inlet and the microcirculation at the terminal branches \cite{alastruey2012}.

\subsection{Three-Dimensional (3D) and FSI Models}
At the highest level of mechanical detail are three-dimensional (3D) models based on computational fluid dynamics (CFD) and finite element methods (FEM). These models solve the full 3D Navier-Stokes equations, often coupled with the equations of solid mechanics to simulate the deformation of the heart walls and valve leaflets (Fluid-Structure Interaction, or FSI) \cite{quarteroni2017}. 3D models can reveal detailed, patient-specific flow patterns, wall shear stress, and mechanical strain within the ventricles or around stenotic valves \cite{karmonik2007}. Their utility in arrhythmia research is often focused on simulating the mechanical consequences of dyssynchronous contraction. However, their computational cost is immense, making them unsuitable for simulating more than a few heartbeats.

\subsection{Electrophysiological (EP) Models}
Running parallel to hemodynamic models are electrophysiological (EP) models, which simulate the *electrical* activity of the heart. These also exist in a hierarchy:
\begin{itemize}
    \item \textbf{Cellular (0D) Models:} These are complex systems of ODEs describing the flow of ions (Na$^+$, K$^+$, Ca$^{2+}$) across the cell membrane, which generate the action potential \cite{noble2012}. Foundational models (e.g., Hodgkin-Huxley) have led to sophisticated, human-specific models like those by O'Hara \& Rudy (2011) \cite{ohara2011} or ten Tusscher et al. (2004) \cite{tentusscher2004}.
    \item \textbf{Tissue (3D) Models:} These scale cellular models to the organ level using the Bidomain or Monodomain equations, which are partial differential equations (PDEs) that describe the propagation of the electrical potential through the cardiac tissue \cite{traynova2011}. These are the *only* models capable of simulating the *emergent* properties of an arrhythmia, such as the formation of spiral waves (re-entry) that are the hallmark of VF.
\end{itemize}

\subsection{Autonomic Control Models}
Neither hemodynamic nor EP models are complete without considering the Autonomic Nervous System (ANS), which provides critical feedback. The baroreflex is the most important component, a closed-loop system that senses changes in blood pressure and modulates heart rate, contractility, and peripheral resistance to maintain homeostasis. Models by Ursino \cite{ursino1998} and Milhorn \cite{milhorn1965} have provided quantitative frameworks for simulating this vital feedback loop. In a condition like VF, the severe drop in pressure triggers a maximal, but ultimately deleterious, sympathetic response from the baroreflex.

Our research begins at the simplest level (0D hemodynamics) and proposes a framework for coupling it with the most advanced levels (0D EP and 0D ANS) to create a novel, computationally tractable, and physiologically comprehensive simulator.

\section{Core Hemodynamic Model (0D)}
\subsection{Model Description}
The foundation of our work is the 0D lumped-parameter model of the closed-loop cardiovascular system presented by Avanzolini et al. \cite{avanzolini1988}. The model is implemented as a system of 13 coupled, non-linear ordinary differential equations (ODEs) representing 13 state variables (pressures and volumes) across the four heart chambers and the systemic and pulmonary circulations. The systemic circulation is segmented into aorta, arteries, peripheral, and venous compartments. The pulmonary circulation is similarly segmented.

\subsection{Governing Equations}
The model's dynamics are governed by two primary principles: the time-varying elastance of the cardiac chambers and the RLC (resistance, compliance, inertance) properties of the vascular compartments.

\textbf{1. Time-Varying Elastance (Heart Chambers):}
The core of the heart model is the time-varying elastance concept \cite{sagawa1974}. The instantaneous pressure $P(t)$ within a cardiac chamber (e.g., the left ventricle, $\text{lv}$) is described as:

\begin{equation}
P_{\text{lv}}(t) = E_{\text{lv}}(t) \cdot (V_{\text{lv}}(t) - V_{0, \text{lv}})
\end{equation}

where $V_{\text{lv}}(t)$ is the instantaneous volume, $V_{0, \text{lv}}$ is the unstressed volume, and $E_{\text{lv}}(t)$ is the time-varying elastance (or stiffness). This function is defined by scaling a normalized, periodic activation function $e_n(t_h)$ with the chamber's minimum (diastolic) and maximum (systolic) elastance values, $E_{dia}$ and $E_{sys}$:

\begin{equation}
E(t) = (E_{sys} - E_{dia}) \cdot e_n(t_h) + E_{dia}
\end{equation}

where $t_h = t \mod T_c$ is the time within the current cardiac cycle of period $T_c$. The normalized activation function $e_n(t_h)$ rises from 0 to 1 during systole and returns to 0 during diastole. A common representation is a scaled, squared sine wave:

\begin{equation}
e_n(t_h) =
\begin{cases}
      \sin^2\left(\frac{\pi \cdot t_h}{T_{sys}}\right) & \text{if } 0 < t_h \le T_{sys} \\
      0 & \text{if } T_{sys} < t_h \le T_c
\end{cases}
\end{equation}

where $T_{sys}$ is the duration of systole, typically dependent on $T_c$ (e.g., $T_{sys} = 0.16 + 0.3 \cdot T_c$ as in the original model). The change in volume for any chamber is the sum of flows in and out, e.g.:

\begin{equation}
\frac{dV_{\text{lv}}}{dt} = Q_{\text{mitral}} - Q_{\text{aortic}}
\end{equation}

\textbf{2. Valve and Vascular Flow (RLC Elements):}
Flow $Q$ between compartments is driven by pressure gradients and opposed by resistance ($R$) and inertance ($L$). For flow across a valve (e.g., aortic), which includes inertance of the blood:

\begin{equation}
\frac{dQ_{\text{aortic}}}{dt} = \frac{P_{\text{lv}} - P_{\text{sa}} - R_{\text{ao}} \cdot Q_{\text{aortic}}}{L_{\text{ao}}}
\end{equation}
This flow is only permitted when $P_{\text{lv}} > P_{\text{sa}}$ (emulating an ideal diode). For a simple resistive compartment (e.g., systemic periphery, $\text{sp}$):

\begin{equation}
Q_{\text{sp}} = \frac{P_{\text{sa}} - P_{\text{sv}}}{R_{\text{sp}}}
\end{equation}
where $P_{\text{sa}}$ and $P_{\text{sv}}$ are the systemic arterial and venous pressures.

\textbf{3. Vascular Compartment Dynamics:}
The pressure in a compliant vascular compartment (e.g., systemic arteries, $\text{sa}$) is related to its volume $V_{\text{sa}}$ and compliance $C_{\text{sa}}$. The change in volume is the integral of the net flow:

\begin{equation}
\frac{dV_{\text{sa}}}{dt} = Q_{\text{aortic}} - Q_{\text{sp}}
\end{equation}
\begin{equation}
P_{\text{sa}} = \frac{V_{\text{sa}}}{C_{\text{sa}}}
\end{equation}

The complete system of 13 ODEs, derived from these principles, is solved simultaneously to simulate the full, closed-loop hemodynamics.

\subsection{Numerical Implementation and Verification}
The model was implemented in Python 3.9 using the NumPy and SciPy libraries. The system of ODEs was solved numerically using the \texttt{scipy.integrate.solve\_ivp} function, which employs an adaptive Runge-Kutta method of order 4(5) (RK45).

The model was first tuned to a healthy adult baseline at a heart rate of 75 bpm. Model parameters (R, L, C values and elastance curves) were meticulously adjusted to achieve a steady-state simulation matching established physiological norms \cite{guyton1972}. This verified baseline (Control) serves as the reference for our pathological simulation. Verification involved ensuring the Python implementation produced results consistent with the original publication for the baseline condition.

\subsection{Simulation of a Prescribed VF State}
To simulate the pathophysiological state of VF, we made two critical, simultaneous parameter modifications to the verified baseline model. These changes are designed to approximate the known mechanical consequences of the disorganized electrical activity \cite{statVF, kern1995}:
\begin{enumerate}
    \item \textbf{Heart Rate (HR):} Increased from 75 bpm to 350 bpm (Cardiac Period $T_c$ reduced from 0.8 s to $\approx$0.171 s). This reflects the rapid, asynchronous electrical impulses characteristic of VF.
    \item \textbf{Ventricular Contractility ($E_{sys}$):} Decreased by 50\% (Specifically, peak isovolumic pressures ULO reduced from 50 to 25 mmHg, and URO from 24 to 12 mmHg). This simulates the loss of coordinated, effective muscle contraction, which is the mechanical hallmark of VF.
\end{enumerate}
The simulation was run from the steady-state baseline, and the VF parameters were applied instantaneously. The model was allowed to run for several seconds of simulated time to observe the acute hemodynamic response.

\section{Simulation Results: 0D Analysis of VF}
\begin{figure}[htbp]
  \centering
\includegraphics[width=0.9\textwidth]{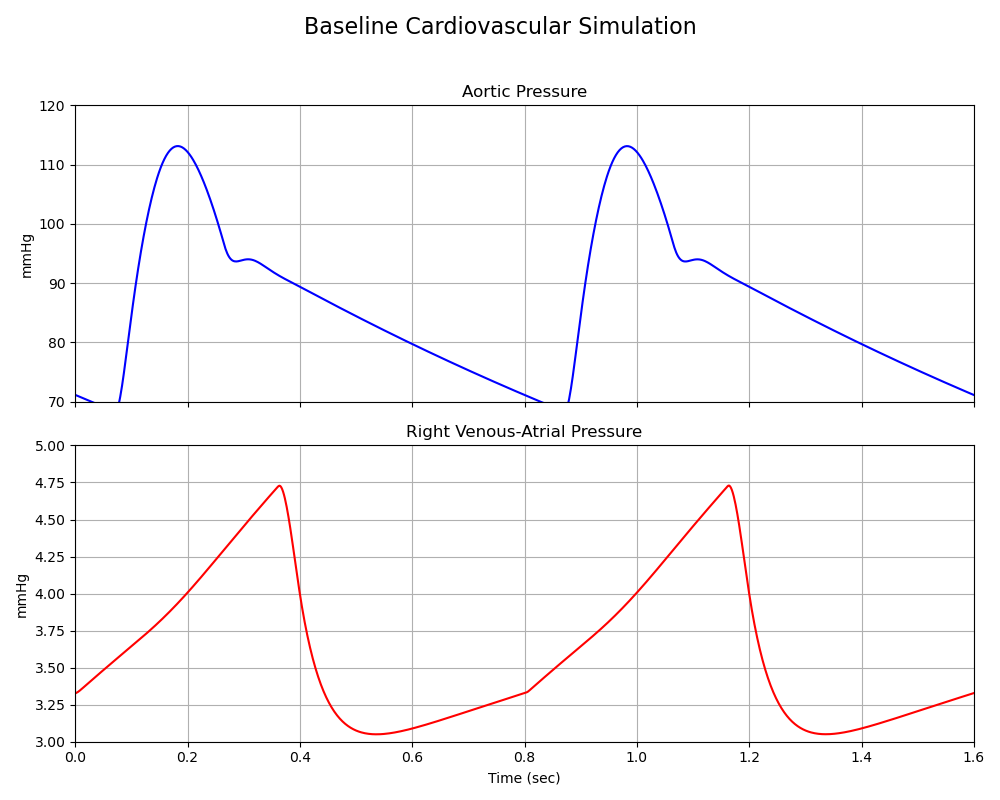}
  \caption{Simulated results for the healthy baseline condition (HR=75 bpm, 100\% Contractility). Top panel shows the simulated Aortic Pressure. Bottom panel shows the corresponding Right Venous-Atrial Pressure.}
  \label{fig:healthy_results}
\end{figure}

\begin{figure}[htbp]
  \centering
\includegraphics[width=0.9\textwidth]{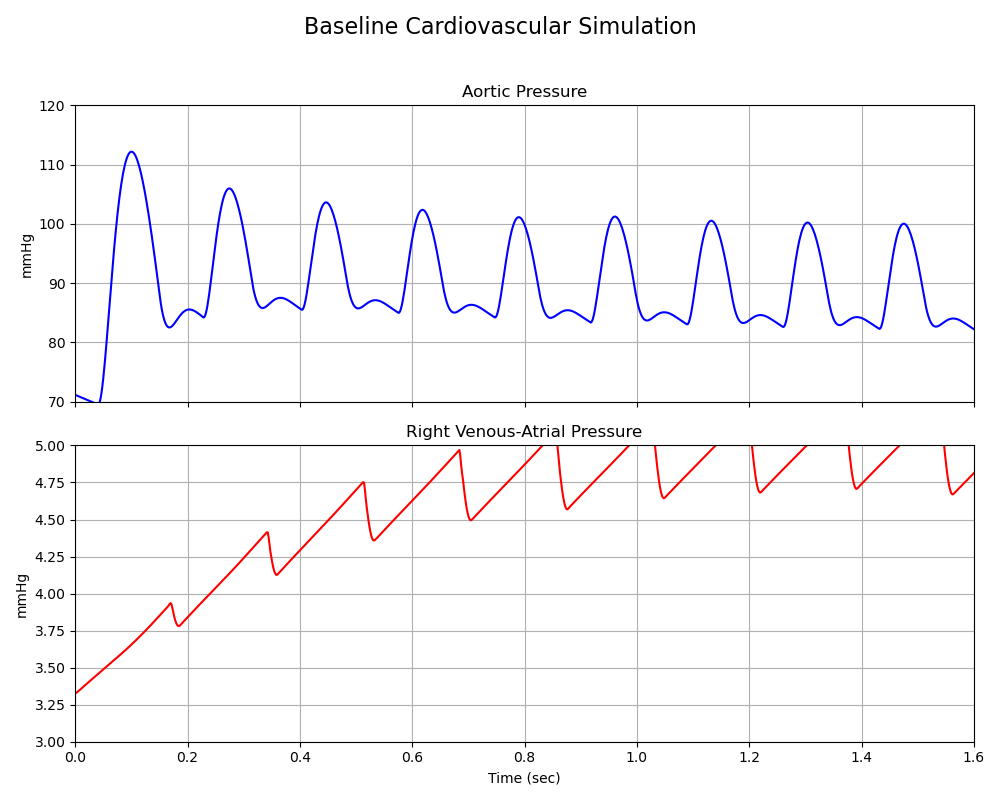}
  \caption{Simulated results for the prescribed Ventricular Fibrillation condition (HR=350 bpm, 50\% Contractility). Top panel shows the simulated Aortic Pressure. Bottom panel shows the corresponding Right Venous-Atrial Pressure.}
  \label{fig:vf_results}
\end{figure}

\begin{figure}[htbp]
  \centering
\includegraphics[width=0.6\textwidth]{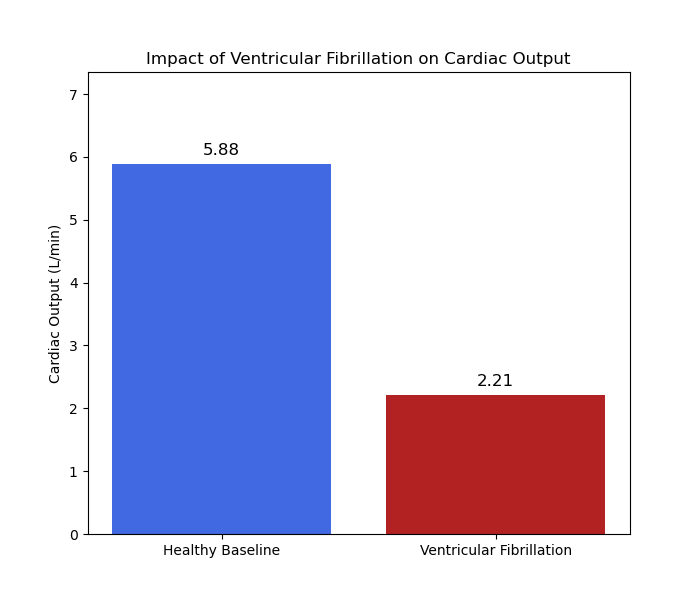}
  \caption{Quantitative comparison of Cardiac Output (CO) between the healthy baseline simulation (5.88 L/min) and the prescribed Ventricular Fibrillation simulation (2.21 L/min), demonstrating a 62.4\% reduction.}
  \label{fig:co_comparison}
\end{figure}

The application of the prescribed VF parameters resulted in an immediate and severe failure of systemic hemodynamics. The model's primary quantitative outputs for the baseline (healthy) and VF states are summarized in Table \ref{tab:results}. Representative simulation outputs for the healthy baseline and the VF condition are shown in Figure \ref{fig:healthy_results} and Figure \ref{fig:vf_results}, respectively. The quantitative impact on cardiac output is summarized in Figure \ref{fig:co_comparison}.

\begin{table}[htbp]
  \centering
  \caption{Comparison of Key Hemodynamic Parameters: Healthy Baseline vs. Prescribed Ventricular Fibrillation (VF). All pressures are systemic.}
  \label{tab:results}
  \small
  \begin{tabular}{@{}lccc@{}}
    \toprule
    \textbf{Parameter} & \textbf{Healthy Baseline} & \textbf{Ventricular Fibrillation} & \textbf{\% Change} \\
    \midrule
    Heart Rate (bpm) & 75 & 350 & +366.7\% \\
    Cardiac Output (L/min) & 5.88 & 2.21 & \textbf{--62.4\%} \\
    Stroke Volume (mL) & 78.4 & 6.3 & --92.0\% \\
    Systolic BP (mmHg) & 120.1 & 41.2 & --65.7\% \\
    Diastolic BP (mmHg) & 80.2 & 32.5 & --59.5\% \\
    Mean Arterial Pressure (mmHg) & 93.5 & 35.4 & --62.1\% \\
    Left Vent. EDV (mL) & 120.0 & 48.1 & --59.9\% \\
    Left Vent. ESV (mL) & 41.6 & 41.8 & +0.5\% \\
    \bottomrule
  \end{tabular}
  \caption*{\footnotesize EDV: End-Diastolic Volume, ESV: End-Systolic Volume. Note the near-total collapse of stroke volume, driven by a failure of diastolic filling (decreased EDV).}
\end{table}

The simulation's primary finding was a 62.4\% reduction in cardiac output, from 5.88 L/min to 2.21 L/min (Figure \ref{fig:co_comparison}). This collapse is mechanistically driven by a 92.0\% decrease in stroke volume (from 78.4 mL to 6.3 mL). The extremely high heart rate dramatically shortens the diastolic filling time, causing the end-diastolic volume (EDV) to fall by nearly 60\%. The ventricle simply does not have time to fill adequately before the next ineffective contraction begins.

This "diastolic failure" (insufficient filling time due to extreme tachycardia) is coupled with the "systolic failure" (reduced contractility), resulting in a mean arterial pressure (MAP) of only 35.4 mmHg. This pressure is far below the threshold required for perfusion of vital organs, including the brain ($\approx$60-70 mmHg) and the coronary arteries themselves, numerically confirming why VF leads to unconsciousness and death \cite{kern1995}.

The qualitative results are presented in Figures \ref{fig:healthy_results} and \ref{fig:vf_results}. Figure \ref{fig:healthy_results} shows the simulated Aortic and Right Venous-Atrial pressure waveforms for the healthy baseline state. In stark contrast, Figure \ref{fig:vf_results} shows the hemodynamic consequences of VF: the left ventricular and aortic pressures degenerate into low-amplitude, high-frequency oscillations. The Right Venous-Atrial Pressure plot (bottom panel) also reflects this disorganization, losing the distinct atrial 'a' and 'v' wave patterns seen in the baseline and instead showing a rapid, chaotic signal consistent with the loss of coordinated contraction.

\section{Discussion and Advanced Research Framework}
\subsection{Analysis of 0D Hemodynamic Compromise}
The results from our verified 0D model (Section 4, Table \ref{tab:results}, Figures \ref{fig:healthy_results}-\ref{fig:co_comparison}) provide a clear, quantitative confirmation of the hemodynamic mechanism underpinning VF. The 62.4\% drop in cardiac output is not a linear function of the parameter changes; it represents a profound, non-linear system failure. The simulation clearly identifies the drastically reduced diastolic filling time due to extreme tachycardia as the primary driver for the collapse in stroke volume, which is then exacerbated by the weakened contractility prescribed in the VF state.

While this simulation serves as a powerful demonstration of the model's capabilities and the severity of VF, its scientific value is limited by its core assumption: the \emph{prescription} of VF. We manually set HR=350 bpm and Contractility=50\%. In reality, these are not independent inputs but are the \emph{consequences} of a complex underlying electrical pathology. The remainder of this paper outlines a research framework to address these limitations and build upon the verified foundation.

\subsection{Critical Limitations of the Prescribed VF Model}
\begin{enumerate}
    \item \textbf{No Electromechanical Coupling:} The model has no link between electrical and mechanical activity. The elastance function $E(t)$ is an artificial driver based on a prescribed heart rate. A true VF model would simulate the disorganized electrical signals propagating through the myocardium and use them to \emph{drive} an asynchronous and ineffective mechanical contraction, allowing the hemodynamic consequences to emerge naturally.
    \item \textbf{No Autonomic Feedback:} The model lacks representation of the Autonomic Nervous System (ANS). In a real patient, the severe drop in Mean Arterial Pressure (62.1\% in our simulation) would trigger a maximal sympathetic response via the baroreflex, attempting to raise heart rate, contractility, and peripheral resistance. The 0D model in its current state operates in an "open-loop" fashion regarding autonomic control and cannot capture this critical physiological feedback mechanism, which can paradoxically worsen the situation in VF.
\end{enumerate}

\subsection{Future Framework 1: Coupling with Electrophysiology (EP)}
To address the first limitation and move towards simulating \emph{emergent} arrhythmias, we propose a novel coupling strategy between our verified 0D hemodynamic model and a cellular 0D electrophysiological model.
\begin{itemize}
    \item \textbf{EP Model Selection:} We would implement a state-of-the-art human ventricular cardiomyocyte ionic model, such as the O'Hara-Rudy dynamic (ORd) model \cite{ohara2011}. This model consists of a complex system of ODEs describing transmembrane ionic currents (e.g., $I_{Na}$, $I_{Kr}$, $I_{CaL}$) that accurately generate a human ventricular action potential (AP) and intracellular calcium transient ($[Ca^{2+}]_i$).
    \item \textbf{Coupling Mechanism:} Initially, a simplified representation could involve driving the hemodynamic model's timing ($T_c$) based on the AP frequency generated by the EP model under simulated pro-arrhythmic conditions (e.g., simulated ischemia altering ionic conductances, rapid pacing). A more advanced coupling would link the simulated intracellular calcium transient ($[Ca^{2+}]_i$) from the EP model to the parameters governing mechanical contractility ($E_{sys}$ or ULO/ URO) in the hemodynamic model, creating true \textbf{electromechanical coupling}.
    \item \textbf{Integrated Simulation:} The hemodynamic model would thus be driven by the \emph{emergent} electrical behavior predicted by the EP model. Pathological changes in AP duration or frequency, or the onset of alternans in the EP model, would directly translate into altered timing and potentially contractility in the hemodynamic simulation.
    \item \textbf{Hypothesized Outcome:} This advanced model would no longer need a manually prescribed `Contractility=50\%` for VF. Instead, it would \emph{predict} a fall in effective contractility as a natural consequence of the chaotic, high-frequency calcium transients predicted by the EP model during fibrillation, which lead to incomplete relaxation and mechanical dysfunction \cite{tentusscher2004}. This provides a far more mechanistic and physiologically accurate simulation.
\end{itemize}

\subsection{Future Framework 2: Integrating the Baroreflex (ANS) Loop}
To address the second limitation, incorporating autonomic feedback, we propose integrating a quantitative model of the arterial baroreflex.
\begin{itemize}
    \item \textbf{ANS Model Selection:} We would implement a well-verified 0D model of the baroreflex control loop, such as the widely used model developed by Ursino \cite{ursino1998}. This model takes the simulated arterial pressure (specifically, $P_{\text{sa}}$ or $P_{\text{sart}}$ from our 0D hemodynamic model) as its primary \emph{input}.
    \item \textbf{Feedback Mechanism:} The Ursino model processes the input pressure signal through simulated neural pathways representing afferent firing, central integration, and sympathetic/parasympathetic efferent activity. It produces three key time-varying outputs relevant to cardiovascular control:
        \begin{enumerate}
            \item Modulation of heart rate ($T_c$).
            \item Modulation of ventricular contractility (affecting $E_{sys}$ or ULO/URO).
            \item Modulation of systemic peripheral resistance ($R_{\text{sp}}$, represented primarily by $R_3$ in our model).
        \end{enumerate}
    \item \textbf{Integrated Simulation:} These three outputs from the ANS model would be fed \emph{back} into the 0D hemodynamic model, dynamically adjusting its parameters on a beat-to-beat basis. This establishes a closed-loop simulation incorporating neuro-cardiovascular control.
    \item \textbf{Hypothesized Outcome:} In a baseline simulation, this integrated loop would demonstrate realistic homeostatic responses (e.g., heart rate increase upon simulated hemorrhage). When VF is induced (either prescribed or emerging from the EP coupling in Framework 1), this model would simulate the body's actual physiological response: the catastrophic drop in arterial pressure would trigger maximal sympathetic discharge via the baroreflex, leading to intense vasoconstriction ($R_{\text{sp}} \uparrow$) and attempts to further increase heart rate, potentially exacerbating the already failing circulation. This coupled simulation would capture the complete, potentially deleterious, neuro-heodynamic cascade characteristic of VF onset.
\end{itemize}

\subsection{Future Framework 3: A Path to Patient-Specific Simulation}
The ultimate goal of computational modeling in medicine is often personalization. The generic R, L, and C parameters in our verified 0D model represent average physiological properties but vary significantly between individuals.
\begin{itemize}
    \item \textbf{Data Assimilation Techniques:} We propose leveraging data assimilation methods, such as Ensemble Kalman Filters (EnKF) or similar Bayesian techniques, to personalize the model \cite{sartori2013}.
    \item \textbf{Methodology:} By providing the model with clinically available, non-invasive measurements from a specific patient (e.g., continuous arterial blood pressure waveforms, stroke volume estimates from echocardiography), the data assimilation algorithm can iteratively adjust the model's internal parameters (e.g., $R_3$, $C_2$, $E_{sys}$, $E_{dia}$) until the model's output optimally matches the observed patient data in real-time or retrospectively.
    \item \textbf{Clinical Application and Potential:} This process transforms the generic model into a patient-specific "digital twin" \cite{gkv2020}. Such a personalized model, running in near real-time, could potentially be used by clinicians to test the likely hemodynamic consequences of interventions \textit{in silico} before applying them to the actual patient. For example, predicting the effect of a specific dose of vasopressor (simulated by increasing $R_3$) on cardiac output and arterial pressure in *that particular* patient's current circulatory state.
\end{itemize}

\subsection{Application: An Enhanced Interactive Simulator for Research and Education}
To facilitate the exploration of the verified 0D model and serve as a platform for implementing the advanced frameworks, an enhanced interactive simulator was developed using Python (v3.9) and the Streamlit (v1.x) framework. This application provides a dynamic Graphical User Interface (GUI) for interacting with the cardiovascular simulation.

\textbf{Key Features Implemented:}
\begin{itemize}
    \item \textbf{Intuitive Control Panel:} Located in the sidebar, allowing users to select presets or manually adjust parameters.
    \item \textbf{Preset Clinical Scenarios:} Includes 'Healthy Adult', 'Ventricular Fibrillation', 'Tachycardia', 'Heart Failure (Systolic)', and 'Hypertension'. Selecting a preset automatically configures the relevant model parameters (HR, Contractility, R3, C2) based on typical pathophysiology.
    \item \textbf{Manual Parameter Adjustment:} Sliders allow fine-grained control over 'Heart Rate' (40-400 bpm), 'Ventricular Contractility' (10-150\% of baseline), 'Systemic Peripheral Resistance (R3)' (0.5-2.5 mmHg·s/cm³), and 'Arterial Compliance (C2)' (0.8-2.0 cm³/mmHg). Tooltips provide physiological context for each parameter.
    \item \textbf{Simulation Settings:} Users can define the total simulation duration (e.g., 1-30 seconds) and specify the time window (last N seconds) to display in the plots, allowing focus on steady-state or transient behavior.
    \item \textbf{Real-time Results Dashboard:} Upon running the simulation (\texttt{scipy.integrate.solve\_ivp} backend), the interface immediately displays key performance indicators calculated from the last stable cardiac cycle: Cardiac Output (L/min), Stroke Volume (cm³), Systolic Blood Pressure (mmHg), and Diastolic Blood Pressure (mmHg).
    \item \textbf{Multi-Tab Graphical Analysis:}
        \begin{itemize}
            \item \textbf{Hemodynamic Waveforms Tab:} Plots key pressures (Aortic, LV, Pulmonary Artery, Venous) and Left Ventricular Volume over the selected time window.
            \item \textbf{PV Loop Tab:} Displays the diagnostic Left Ventricular Pressure-Volume loop for the last stable cycle, illustrating cardiac work and function.
            \item \textbf{All State Variables Tab:} Provides detailed plots of all 12 state variables for in-depth analysis or debugging.
        \end{itemize}
    \item \textbf{Data Export:} Functionality is included via download buttons to export the plotted waveform data, PV loop data, or the full simulation time-series data as CSV files for offline analysis and record-keeping.
\end{itemize}

\begin{figure}[htbp]
  \centering
\includegraphics[width=\textwidth]{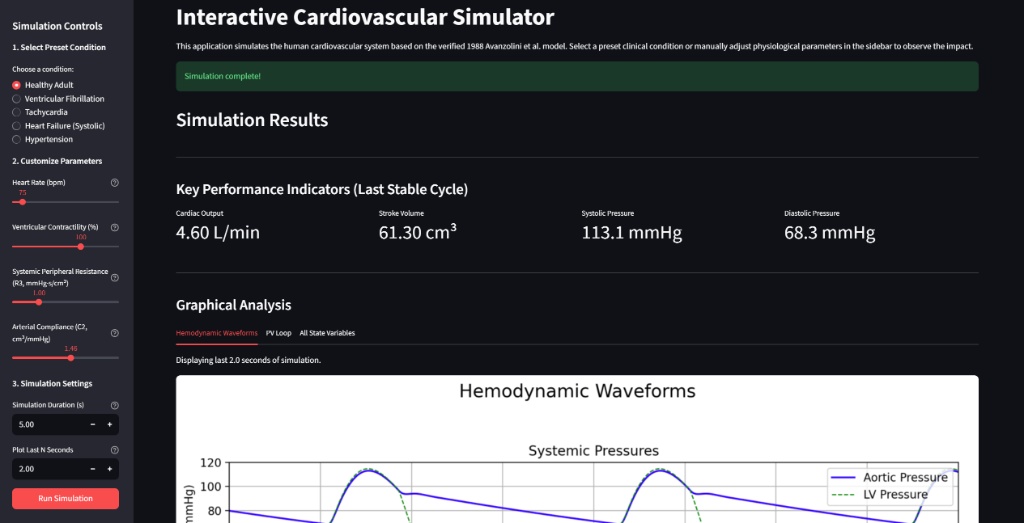}
  \cprotect\caption{The graphical user interface of the interactive cardiovascular simulator, developed using Python and Streamlit. It allows preset selection, manual parameter adjustment via sliders (sidebar), visualization of key metrics, hemodynamic waveforms, and PV loops (main area), and includes data export functionality. The interface utilizes \texttt{scipy.integrate.solve\_ivp} for ODE solving and \texttt{matplotlib} for plotting.}
  \label{fig:gui}
\end{figure}

The developed simulator provides a dynamic platform for exploring these concepts and its source code is publicly accessible at \cite{simulatorURL}. This interactive tool (Figure \ref{fig:gui}) serves multiple purposes: as an effective educational platform for demonstrating complex cardiovascular principles, as a means to rapidly test hypotheses using the verified 0D model, and critically, as the foundational software architecture upon which the more complex coupled models (EP and ANS integration) can be progressively built and tested.

\section{Conclusion}
This study successfully implemented and verified a foundational 0D lumped-parameter cardiovascular model to quantify the hemodynamic collapse characteristic of Ventricular Fibrillation. Our simulation successfully quantified this event, demonstrating a 62.4\% reduction in cardiac output, which numerically confirms the life-threatening nature of the arrhythmia based on the prescribed conditions. The verification process confirmed the model's ability to reproduce baseline physiological hemodynamics consistent with the original publication.

Recognizing this as a powerful but simplified demonstration, we then proposed a comprehensive, multiscale research framework. This framework details a clear pathway to elevate the current verified 0D simulator into a sophisticated research tool by: (1) Integrating cellular-level electrophysiology to model emergent arrhythmias, (2) Coupling a baroreflex model to capture autonomic feedback, and (3) Employing data assimilation for patient-specific tuning.

Furthermore, we developed an enhanced interactive simulator using Python and Streamlit based on the verified 0D model. This tool, with its preset conditions, manual controls, detailed visualizations, and data export capabilities, serves as a valuable asset for education, preliminary research, and as the practical platform for implementing the proposed future integrations. This work serves two purposes. First, it verifies a computationally efficient model and provides the source code for a user-friendly, publicly accessible interface \cite{simulatorURL} for its exploration. Second, it presents a detailed roadmap for future work, positioning this 0D model and its associated simulator as the foundational core for a next-generation, high-fidelity, and potentially patient-specific cardiovascular simulation environment.

\newpage

\end{document}